\documentclass[twocolumn,prl,superscriptaddress,noshowpacs,epsf]{revtex4}

\usepackage{picins,graphicx}
\usepackage{boxedminipage}
\usepackage{amsmath}

\begin{document}

\title{Atomic physics and quantum optics using superconducting
circuits}

%
\author{J. Q. You}
\affiliation{Department of Physics, State Key Laboratory of Surface
Physics, and Key Laboratory of Micro and Nano Photonic Structures
(Ministry of Education), Fudan University, Shanghai 200433, China}
\affiliation{Advanced Science Institute, RIKEN, Wako-shi 351-0198,
Japan}
\author{Franco Nori}
\affiliation{Advanced Science Institute, RIKEN, Wako-shi 351-0198,
Japan} \affiliation{Physics Department, The University of Michigan,
Ann Arbor, MI 48109-1040, USA}

\begin{abstract}
Superconducting circuits based on Josephson junctions exhibit
macroscopic quantum coherence and can behave like artificial atoms.
Recent technological advances have made it possible to implement
atomic-physics and quantum-optics experiments on a chip using these
artificial atoms. This review presents a brief overview of the
progress achieved so far in this rapidly advancing field. We not
only discuss phenomena analogous to those in atomic physics and
quantum optics with natural atoms, but also highlight those not
occurring in natural atoms. In addition, we summarize several
prospective directions in this emerging interdisciplinary field.
\end{abstract}

\maketitle

Superconducting circuits with Josephson junctions can behave as
artificial atoms. In these quantum circuits, the Josephson junctions
act as nonlinear circuit elements (see Box~1). Such nonlinearity in
a circuit ensures an unequal spacing between energy levels, so that
the lowest levels can be individually addressed by using external
fields (see, e.g.,
\cite{NEC99,Wal,Saclay,Han,Martinis,rev1,rev2,rev4,rev5}).
Experimentally, these circuits are fabricated on a micrometer scale
and operated at mK temperatures. Because of the reduced
dimensionality and thanks to the superconductivity, the
environment-induced dissipation and noise are greatly suppressed, so
the circuits can behave quantum mechanically.

Superconducting circuits based on Josephson junctions have recently
become subjects of intense research because they can be used as
qubits---controllable quantum two-level system---for quantum
computing (see, e.g., \cite{rev1,rev2,rev4,rev5} for reviews). Even
though the typical decoherence times of these circuits fall short of
the requirements for quantum computation, their macroscopic quantum
coherence is sufficient for them to exhibit striking quantum
behaviors. These circuits can have a number of superconducting
eigenstates with discrete eigenvalues lower than the energy levels
of the quasi-particle excitations that involve breaking Cooper
pairs. This property allows these circuits to behave like
superconducting artificial atoms.
Indeed, there is a deep analogy between natural atoms and the
artificial atoms made from superconducting circuits (see Box~2).
Both have discrete energy levels and can exhibit coherent quantum
oscillations between these levels. Whereas natural atoms may be
controlled using visible or microwave photons that excite electrons
from one state to another, the artificial atoms in the circuits are
driven by currents, voltages and microwave photons that excite the
system from one macroscopic quantum state to another.

Differences between superconducting circuits and natural atoms
include the different energy scales in the two systems, and how
strongly each system couples to its environment; the coupling is
weak for atoms and strong for circuits. In contrast to naturally
occurring atoms, artificial atoms can be designed with specific
characteristics and fabricated on a chip using standard
lithographical technologies. With a view to applications, this
degree of tunability is an important advantage over natural atoms.
Thus, in a controllable manner, superconducting circuits can be used
to test fundamental quantum mechanical principles at a macroscopic
scale, as well as to demonstrate atomic physics and quantum optics
on a chip. Moreover, artificial atoms can be designed to have exotic
properties that do not occur in natural atoms.

\vspace{.01cm}\noindent
\begin{boxedminipage}[t]{3.4in}

{\bf\large Box~1. The Josephson junction as a nonlinear inductor}

\begin{center}
\parpic{}{
\includegraphics[width=1.7in,bbllx=0,bblly=0,bburx=329,bbury=124]
{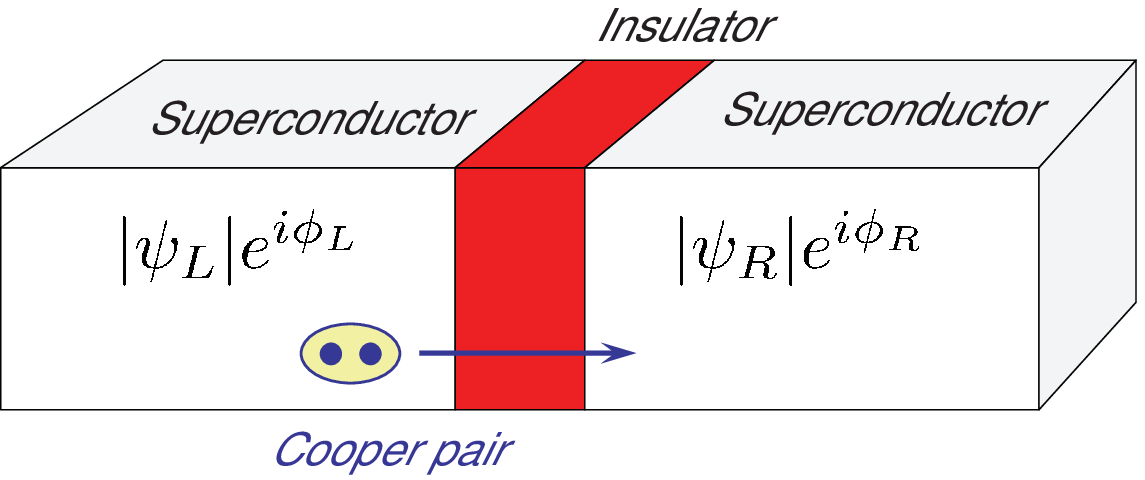} }
\end{center}

A superconductor contains many paired electrons, called Cooper
pairs, which condense into the same macroscopic quantum state
described by the wavefunction $|\psi|e^{i\phi}$, with $|\psi|^2$
being the density of Cooper pairs. In the absence of applied
currents or magnetic fields, the phase $\phi$ is the same for all
Cooper pairs. A Josephson junction is composed of two bulk
superconductors separated by a thin insulating layer through which
Cooper pairs can tunnel (see the figure above).
The supercurrent through the junction is
$I=I_c\sin\varphi$,
where the critical current $I_c$ is related to the Josephson
coupling energy $E_J$ of the junction by $I_c=(2e/\hbar)E_J$,
and $\varphi=\phi_L-\phi_R$ is the phase difference of the two
superconductors across the junction. The time variation of this
phase difference is related to the potential difference $V$ between
the two superconductors:
$\dot{\varphi}=\left(\frac{2\pi}{\Phi_0}\right)V$,
where $\Phi_0=h/2e$ is the magnetic-flux quantum. From the
definition of the inductance $V=L_J\dot{I}$,  it follows that
$L_J=\frac{\Phi_0}{2\pi I_c\cos\varphi}$,
indicating that the Josephson junction behaves like a nonlinear
inductor.


\end{boxedminipage}


\vspace{.01cm}\noindent
\begin{boxedminipage}[t]{3.4in}

{\bf\large Box~2. Artificial and natural atoms}

\begin{center}
\parpic{}{
\includegraphics[width=2.6in,bbllx=0,bblly=0,bburx=461,bbury=334]
{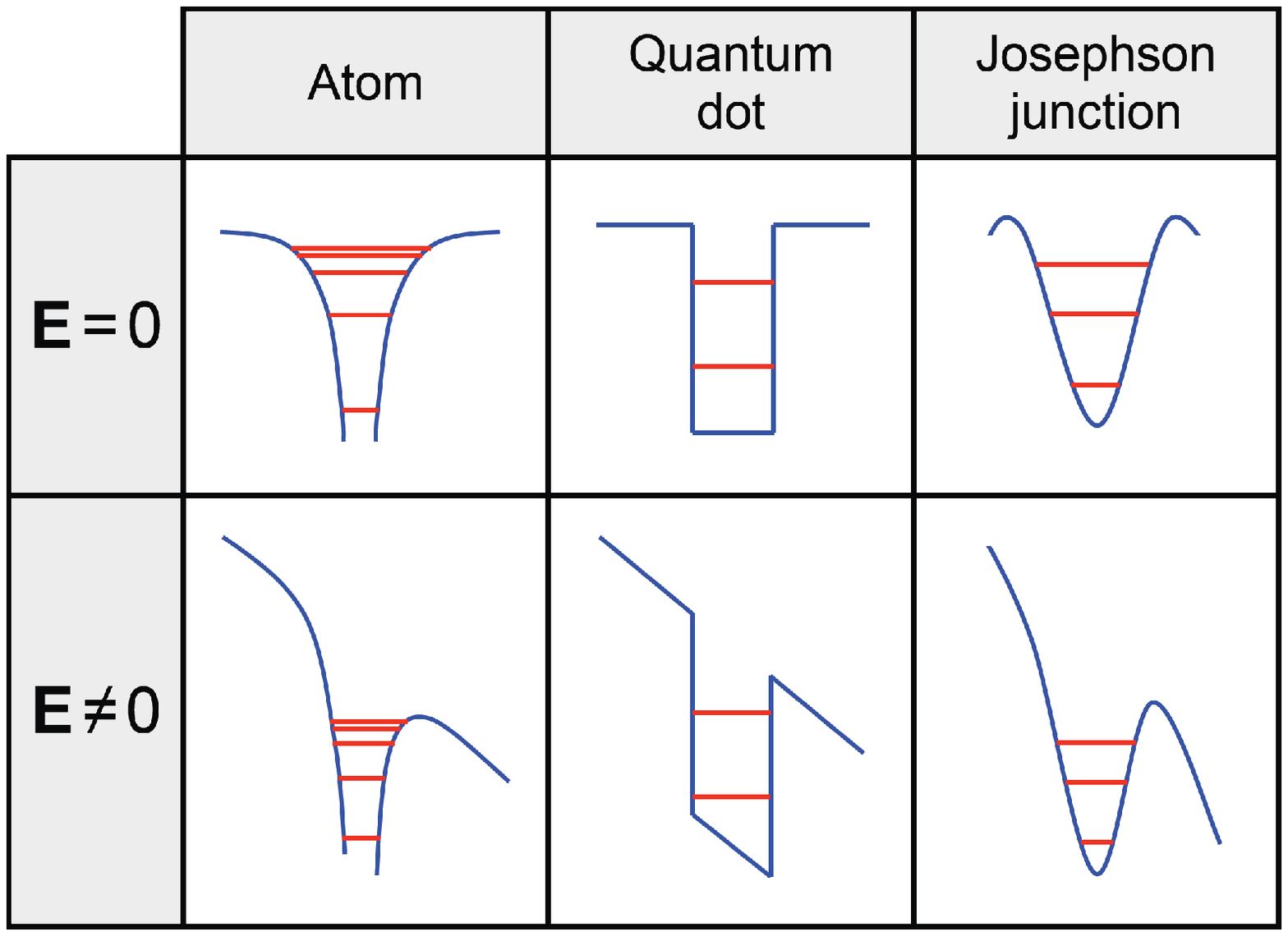} }
\end{center}

Potential energy (in blue) and discrete energy levels (in red) for
an atom, a quantum dot and a Josephson junction in the absence
(${\bf E}=0$) and presence (${\bf E}\neq 0$) of an externally
applied electric field, respectively (see the figure above). Owing
to their confinement, the electrons in the atom and the quantum dot
have discrete energy levels. The Cooper pairs confined in the
potential well of the Josephson coupling energy also have discrete
energy levels and the junction can be regarded as a superconducting
artificial atom.

\end{boxedminipage}
\vspace{.15cm}\noindent


\vspace{.2cm}\noindent

In this review, we highlight the atomic-physics and quantum-optics
phenomena found in superconducting circuits.
The novel physics in these artificial atoms will be emphasized,
including phenomena that do not occur in natural atoms. We also
summarize several prospective directions in this emerging
interdisciplinary field. Some of the examples in this brief overview
relate to our work, because we are more familiar with them.

\vspace{.8cm}\noindent

{\bf\large Superconducting circuits as artificial atoms}
\vspace{.3cm}\noindent

\begin{figure*}
\includegraphics[width=5in,bbllx=25,bblly=345,bburx=570,bbury=742]
{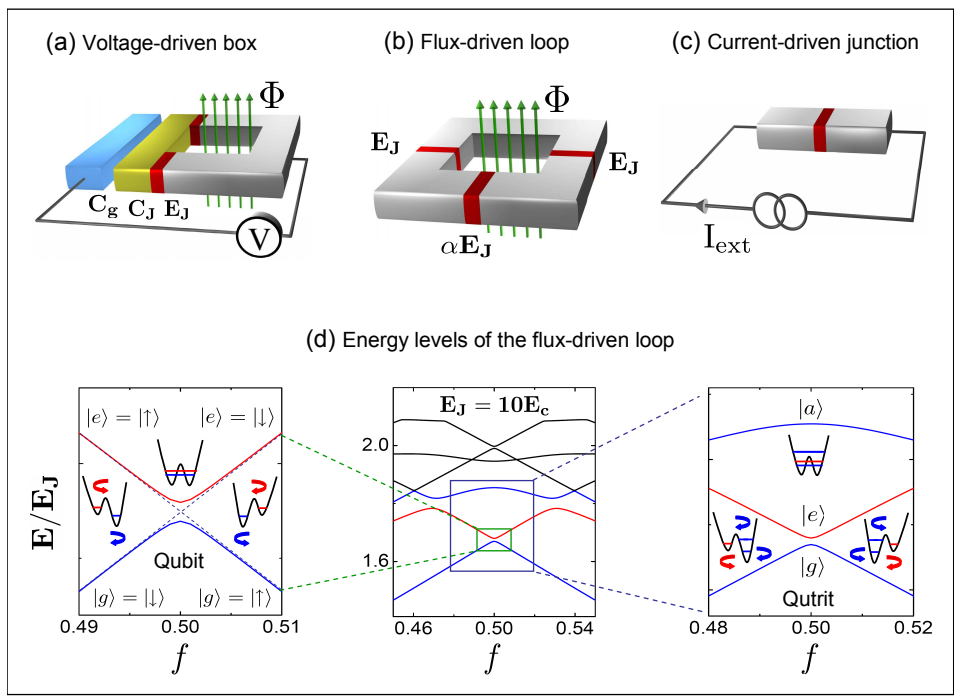}
\caption{{\bf Superconducting circuits as artificial atoms.} {\bf
(a)}~A Cooper-pair box biased by a gate voltage $V_g$ and
implemented in the charge regime $E_J/E_c\ll 1$.
The SQUID loop provides an effective Josephson coupling energy tuned
by the threading magnetic flux $\Phi$. See main text for
nomenclature.
{\bf (b)}~A superconducting loop interrupted by three Josephson
junctions and implemented in the phase regime $E_J/E_c\gg 1$. The
two identical Josephson junctions have coupling energy $E_J$ and
capacitance $C$, while both the Josephson coupling energy and the
capacitance of the smaller junction are reduced by a factor
$\alpha$, where $0.5<\alpha<1$. The three-junction loop is biased by
a flux $\Phi$ such that $f\equiv\Phi/\Phi_0\sim\frac{1}{2}$.
{\bf (c)}~A Josephson junction biased by a current $I_{\rm ext}$,
which is also implemented in the phase regime and has a much larger
ratio $E_J/E_c$. {\bf (d)}~Energy levels of the flux-driven
three-junction loop (the curves in the middle panel). With the
lowest two energy levels involved (the curves in the left panel,
which are enlarged from the smaller rectangle in the middle panel),
the flux-driven loop can behave like a coherent and controllable
quantum two-level system (qubit), while the circuit can behave like
a coherent and controllable three-level system (qutrit) when using
the lowest three levels (the curves in the right panel, which are
enlarged from the larger rectangle in the middle panel). Moreover,
in the left, top and right insets of the left (right) panel, the two
(three) energy levels are also displayed in the double potential
well for $f<\frac{1}{2}$, $f=\frac{1}{2}$ and $f>\frac{1}{2}$, where
the clockwise and anticlockwise arrows represent the circulating
supercurrent states in the flux-driven three-junction loop. In {\bf
(a)-(c)}, the qubit employs the charge states on a single island,
the persistent-current states in a double potential well, and the
anharmonic-oscillator states in a single potential well,
respectively.
Also, a flux-driven superconducting loop with a different number of
Josephson junctions, e.g., one~\cite{Lukens} or four
junctions~\cite{Bert}, can be used for a flux qubit. Furthermore, a
flux-driven single-junction loop can be used as a phase qubit when
working with the energy levels in a tilted potential
well~\cite{Simmonds-PRL04}, as in a current-driven junction.}
\label{fig1}
\end{figure*}

Two important energy scales determine the quantum mechanical
behavior of a Josephson-junction circuit: The Josephson coupling
energy $E_J$ and the electrostatic Coulomb energy $E_c=(2e)^2/2C$
for a single Cooper pair, where $e$ is the electronic charge, and
$C$ is either the capacitance $C_J$ of a Josephson junction or the
capacitance of a superconducting island called a Cooper-pair box
(i.e., the sum of the gate capacitance $C_g$ and the relevant
junction capacitance), depending on the circuit. Figure~\ref{fig1}
summarizes three kinds of superconducting circuits implemented in
different regimes of $E_J/E_c$; Fig.~1(a) shows the voltage-driven
box (also known as a Cooper-pair box) for a charge
qubit~\cite{NEC99}, Fig.~1(b) the flux-driven three-junction loop
for a flux qubit~\cite{Wal} and Fig.~1(c) the current-driven
junction for a phase qubit~\cite{Han,Martinis}. As a typical
example, energy levels of the flux qubit are shown in Fig.~1(d).
Moreover, hybrid superconducting qubits are possible. For instance,
a Cooper-pair box can behave like a charge-flux qubit~\cite{Saclay}
when $E_J/E_c\sim 1$. As for the flux qubit, by reducing the ratio
$E_J/E_c$, the charge noise can become dominant over the flux
noise~\cite{YHAN} and the circuit behaves more like a charge qubit.
In this circuit, when $\alpha<0.5$ (here $\alpha$ is the ratio of
the Josephson coupling energy between the smaller and larger
junctions in the loop), the double-well potential converts to a
single-well potential and the circuit behaves like a phase
qubit~\cite{YHAN,Steffen-PRL10}. One can shunt a large capacitance
to the small junction~\cite{YHAN,Steffen-PRL10} to suppress the
charge noise in this circuit. Also, this large capacitance shunted
to the Josephson junction can be used to reduce the charge noise in
the Cooper-pair box~\cite{Koch}, so as to implement the circuit in
the phase regime. Below we highlight several aspects of the
atomic-physics and quantum-optics phenomena found in superconducting
circuits.

\vspace{.4cm}\noindent

{\bf Cavity quantum electrodynamics.} A quantized electromagnetic
field can coherently exchange energy with a two-level system,
usually in a tiny (micrometer-scale) cavity. This energy exchange
process involves a fundamental phenomenon called Rabi oscillation;
the two-level system and the field exchange a quantum of energy back
and forth at a characteristic frequency known as the Rabi frequency.
When the field is in resonance with the system, the Rabi frequency
is proportional to the system-field coupling strength. The most
elementary of such coherent processes involves the interaction of a
two-level system with a single photon in the cavity. The exchange of
energy between the system and the single photon is observable
when the Rabi frequency is larger than the decay rates of the
two-level system and the cavity. This photon-atom coupling has been
achieved for a variety of atoms interacting with the light field in
a cavity and forms the basis of cavity quantum electrodynamics
(QED).
Cavity QED with superconducting circuits was
proposed~\cite{You-Nori, Blais} and experimentally
achieved~\cite{CQED1,resonator} in systems where superconducting
qubits were employed as two-level artificial atoms. For the cavity,
a single-mode $LC$ resonator~\cite{CQED1} and a multi-mode coplanar
waveguide resonator~\cite{resonator} have been used. Significantly,
the strong-coupling limit for a superconducting qubit in a cavity
can be attained much more easily than for a natural atom in a
cavity~\cite{rev5,Blais} by suitably designing the system
parameters. It allows, e.g., the observation of the Lamb shift for a
superconducting qubit in the cavity~\cite{Fragner-Science}.
Moreover, both the superconducting qubit and the cavity can be
fabricated on the same chip.
For a review on cavity QED with superconducting qubits, see, e.g.,
\cite{rev5}.

Because they can be designed with specific system parameters,
superconducting quantum circuits are suited for achieving the
so-called ultrastrong-coupling regime, where the qubit-photon
coupling strength is comparable to the energy scales of the qubit
and the photon~\cite{Devoret}. Indeed, there have been a number of
theoretical studies of this system, analyzing some of its rich
static and dynamical properties (see, e.g.,
\cite{ultra4,ultra6,ultra7}). Also, the ultrastrong coupling between
a superconducting flux qubit and a coplanar waveguide~\cite{Solano1}
or an $LC$ resonator~\cite{Solano2} has recently been demonstrated
in experiments. One can expect to find new phenomena in this
ultrastrong-coupling regime which are not present in the
conventional weak- and moderately strong-coupling regimes. In
addition, dressed states of a superconducting charge qubit and an
intense microwave field  were experimentally observed by embedding
the circuit in an $LC$ oscillator~\cite{Wilson}. The tunability of
these dressed states allows one to explore both resonant and
dispersive coupling regimes.

\begin{figure*}
\includegraphics[width=4.1in,bbllx=40,bblly=277,bburx=560,bbury=633]
{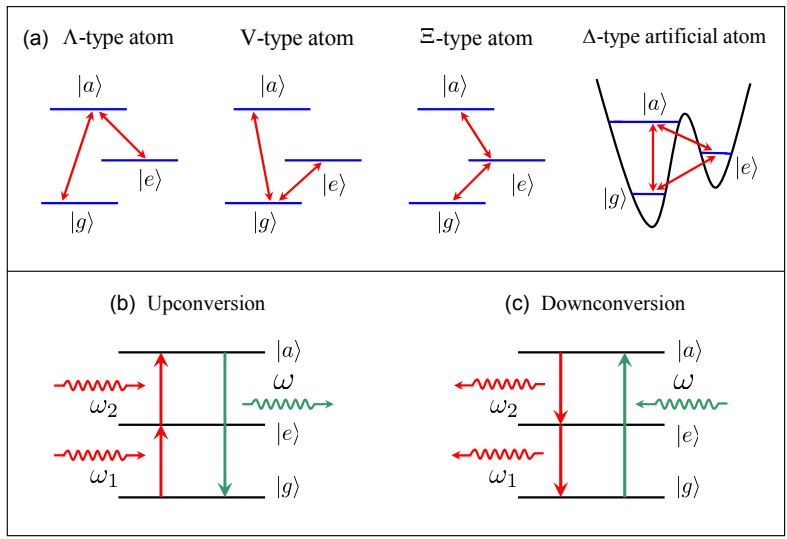}\caption{{\bf Three-level atoms and frequency conversion.}
{\bf (a)}~Energy levels of natural atoms of the $\Lambda$, $V$, and
$\Xi$ types, as well as a $\Delta$-type artificial atom consisting
of a flux-driven three-junction loop. The allowed dipole transitions
between energy levels are indicated in red. In contrast to naturally
occurring atoms, the three dipole transitions among the states
$|g\rangle$, $|e\rangle$ and $|a\rangle$ are all allowed in the
$\Delta$-type three-level artificial atom. Here $|g\rangle$ and
$|e\rangle$ denote the ground and first excited states, while
$|a\rangle$ denotes either the second or another excited state. {\bf
(b)}~Frequency upconversion in a $\Delta$-type artificial atom. Here
$\omega_1=(E_e-E_g)/\hbar$, $\omega_2=(E_a-E_e)/\hbar$, and
$\omega=(E_a-E_g)/\hbar$, with $E_i$ ($i=g,e$ or $a$) being the
energy level of the state $|i\rangle$. When two microwave photons,
one with frequency $\omega_1$ and the other with $\omega_2$, are
successively absorbed by the artificial atom, it can emit a
microwave photon with frequency $\omega=\omega_1+\omega_2$ via the
transition $|a\rangle\rightarrow|g\rangle$. {\bf (c)}~Frequency
downconversion in the $\Delta$-type artificial atom. When a
microwave photon with frequency $\omega$ is absorbed by the
artificial atom, the sequential transitions
$|a\rangle\rightarrow|e\rangle$ and $|e\rangle\rightarrow|g\rangle$
can produce two microwave photons with frequencies $\omega_1$ and
$\omega_2$, respectively. In particular, when $E_a-E_e=E_e-E_g$, the
upconversion in {\bf (b)} converts two photons with frequency
$\frac{1}{2}\omega$ to one photon with frequeny $\omega$, while the
downconversion in {\bf (c)} converts one photon with frequency
$\omega$ to two photons with frequeny $\frac{1}{2}\omega$. Natural
atoms cannot perform up- or down-conversion, unless aided by
nonlinear effects. However, artificial atoms can.}
\label{fig2}
\end{figure*}

\vspace{.3cm}\noindent

{\bf Selecting quantum transitions.}  In natural atoms, the
electronic state at each orbital level has a well-defined parity
symmetry, either even or odd. Under the dipole approximation, the
interaction Hamiltonian between the atom and the time-dependent
electric field has an odd parity. Thus, to have a nonzero dipole
transition matrix element, there should be a parity change between
the initial and final states, in addition to the constraints on
azimuthal and magnetic quantum numbers of the electronic states.
This optical selection rule dictates that only three types of
three-level systems, called $\Lambda$-, $V$-, and $\Xi$-type atoms
[see Fig.~\ref{fig2}(a)], exist for the natural atoms, where no
dipole transition between electronic states with the same parity are
allowed. However, selection rules can be different for
superconducting artificial atoms. For instance, in the dipole
approximation, the interaction Hamiltonian between a flux qubit
circuit and a time-dependent magnetic field does not have a
well-defined parity, except at the point with a static magnetic flux
$f\equiv\Phi/\Phi_0=\frac{1}{2}$, where $\Phi$ is the static
magnetic flux applied to the loop and $\Phi_0$ the magnetic-flux
quantum. At this particular point, the interaction Hamiltonian has
an odd parity. Owing to the parity symmetries of the artificial-atom
states at $f=\frac{1}{2}$, the lowest three levels of the circuit
behave like a $\Xi$-type or ladder-type artificial
atom~\cite{Liu-PRL}. In this case, the dipole transition between
$|g\rangle$ and $|a\rangle$ is forbidden, while the other two
transitions (among states $|g\rangle$, $|e\rangle$ and $|a\rangle$)
are allowed [see Fig.~2(a) for nomenclature]. However, when
$f\neq\frac{1}{2}$, the parity symmetry is broken for the
interaction Hamiltonian. Therefore, all three dipole transitions
among $|g\rangle$, $|e\rangle$ and $|a\rangle$ are possible,
allowing the atom to be $\Delta$-type because of the triangle-shaped
transitions among the three energy levels. Now (when $f\neq
\frac{1}{2}$) the superconducting circuit behaves as a $\Delta$-type
cyclic artificial atom, where one- and two-photon processes can
coexist~\cite{Liu-PRL}.

This $\Delta$-type artificial atom can be used for the upconversion
and downconversion of the photon frequency [see Figs.~\ref{fig2}(b)
and \ref{fig2}(c)]. In these frequency conversions, all transitions
involve only linear processes; this is in sharp contrast to the
conventional frequency conversion in nonlinear optics, where a
nonlinear medium is used and the nonlinear effect facilitates
converting the frequency of the photons. Recently, the frequency
upconversion of a microwave photon was experimentally demonstrated
in a flux qubit~\cite{Deppe}. This experiment explained the observed
coexistence of one- and two-photon processes as due to the
symmetry-breaking of the system Hamiltonian, when varying the
applied magnetic flux away from $f=\frac{1}{2}$.

In the experiment reported in \cite{Groot}, two microwave fields
were applied simultaneously to a superconducting circuit containing
two coupled flux qubits. The interference between the processes that
correspond to a selected excitation by the applied microwave fields
can be controlled and used to activate or suppress a given
transition. Thus, this method effectively creates artificial and
controllable selection rules.

\vspace{.4cm}\noindent

\begin{figure}
\includegraphics[width=3.1in,bbllx=97,bblly=298,bburx=475,bbury=623]
{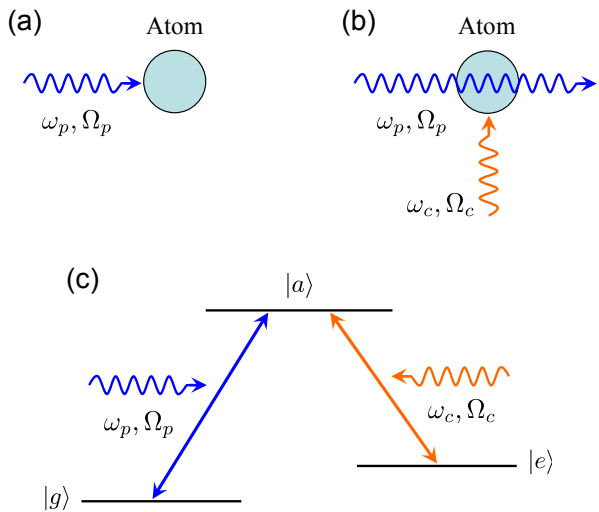} \caption{{\bf Electromagnetically induced transparency.}
{\bf (a)}~A probe light field is absorbed by natural or artificial
atoms when the frequency of the light field is resonant with a
particular separation between two atomic energy levels. {\bf
(b)}~However, the probe light field can go through the natural or
artificial atoms when a suitable control light field also drives the
atoms. {\bf (c)}~$\Lambda$-type three-level atomic system for EIT.
The frequency $\omega_p$ of the probe light field is resonant with
the energy separation between states $|a\rangle$ and $|g\rangle$,
and the frequency $\omega_c$ of the control light field is resonant
with the energy separation between states $|a\rangle$ and
$|e\rangle$. The Rabi frequency $\Omega_p$ ($\Omega_c$) quantifies
the coupling strength between the probe (control) light field and
the atom.} \label{fig3}
\end{figure}

{\bf Electromagnetically induced transparency.} Quantum interference
can be introduced to control the propagation of light through an
atomic medium consisting of three-level atoms or qutrits. We now
consider a $\Lambda$-type three-level atom; to control the
propagation of a probe light field in resonance with the dipole
transition $|a\rangle\leftrightarrow|g\rangle$ (that is, at the
probe frequency $\omega_p$), one can drive the atom by a second
(control) field that is in resonance with the transition
$|a\rangle\leftrightarrow|e\rangle$ at the control frequency
$\omega_c$ (see Fig.~\ref{fig3}). Now the amplitude of the
wavefunction for the state $|a\rangle$ is driven by two terms; one
proportional to the probability amplitude of the ground state
$|g\rangle$ and the other proportional to the probability amplitude
of state $|e\rangle$. The combined effect of these two fields is to
pump the atom into a coherent superposition of the states
$|g\rangle$ and $|e\rangle$ known as a dark state. In such a case,
the two driving terms can interfere and cancel each other, i.e.,
they have equal magnitudes but opposite signs. Under this
destructive quantum interference, the occupation probability at
state $|a\rangle$ is zero, leading to a vanishing light absorption
even in the presence of light fields. This effect is called
electromagnetically induced transparency (EIT) in quantum
optics~\cite{Harris,Scully} and also applies to $V$- and $\Xi$-type
atoms. This important effect has a variety of applications. For
instance, a medium with EIT can slow down and even stop or trap
light.

We now consider EIT in a more quantitative manner. Let $\Omega_p$
($\Omega_c$) be the Rabi frequency that quantifies the coupling
strength between the probe (control) light and the atom. Here we
assume that a $\Lambda$-type atom can be prepared in the initial
state
$|\Psi(0)\rangle=(\Omega_c|g\rangle-\Omega_p|e\rangle)/\Omega$,
where $\Omega=\sqrt{\Omega_p^2+\Omega_c^2}$. When the EIT occurs,
the atom will be trapped in this dark state, i.e.,
$|\Psi(t)\rangle=(\Omega_c|g\rangle-\Omega_p|e\rangle)/\Omega$, for
a time which is dependent on the decoherence rate of the atom.
Usually, it is not easy for an atom to be prepared in the initial
state $|\Psi(0)\rangle$, when $\Omega_p$ is comparable to
$\Omega_c$. Instead, the atom can be naturally prepared in the
ground state $|g\rangle$. If strong control and weak probe fields
are chosen so that $\Omega_c\gg\Omega_p$, the dark state
$|\Psi(t)\rangle$ is close to the initial state $|g\rangle$. In such
a case, the combined action of the control and probe fields can
easily drive the atom from the ground state into the dark
state~\cite{Scully}. This is the reason why a strong control field
and a weak probe field are used to experimentally implement EIT in
an atomic medium.

At $f=\frac{1}{2}$, the flux qubit circuit can behave like a
$\Xi$-type artificial atom~\cite{Liu-PRL,maser}. The circuit can
also behave approximately like a $\Lambda$-type artificial atom when
$f\neq\frac{1}{2}$, if the dipole transition rate between
$|g\rangle$ and $|e\rangle$ is much smaller than the rates for the
other two transitions~\cite{maser}. As discussed above, these
selection rules are related to the parity symmetries of the
artificial-atom states. In contrast to natural atoms, the tunability
of a superconducting circuit can be used to prepare the artificial
atom in an initial state that is close to the dark state
$|\Psi\rangle$ with arbitrary $\Omega_c$ and $\Omega_p$. Therefore,
it is experimentally feasible to produce EIT in a single artificial
atom for either strong or weak control and probe fields. This is an
important advantage of superconducting circuits compared to natural
atoms. EIT using superconducting circuits has been studied
theoretically (e.g., \cite{Orlando,Dutton,Ian}) and
experimentally~\cite{Sill,Abdu}. In ~\cite{Abdu}, this phenomenon
was experimentally shown using a four-junction loop biased at
$f=\frac{1}{2}$, where the circuit behaves as a $\Xi$-type
artificial atom.

\vspace{.4cm}\noindent

{\bf State population inversion and lasing.} A laser is composed of
an amplifying medium inside a resonant optical cavity. When the
system is driven, a state population inversion (SPI) can be achieved
for the atoms or molecules in the amplifying medium. Moreover, there
is a positive feedback between the emitted light and the amplifying
medium. Because of this positive feedback and the nature of the
stimulated photon emission, the laser has a large net optical gain
and the emitted photons have the same direction, phase and
polarization. These advantages mean that the laser has a variety of
applications in different fields. Recently, several studies have
considered lasing using only a single artificial atom, both
theoretically~\cite{maser,Armour,Hauss,Ashhab} and
experimentally~\cite{lasing,Grajcar}.

\begin{figure}
\includegraphics[width=2.7in,bbllx=103,bblly=362,bburx=316,bbury=606]
{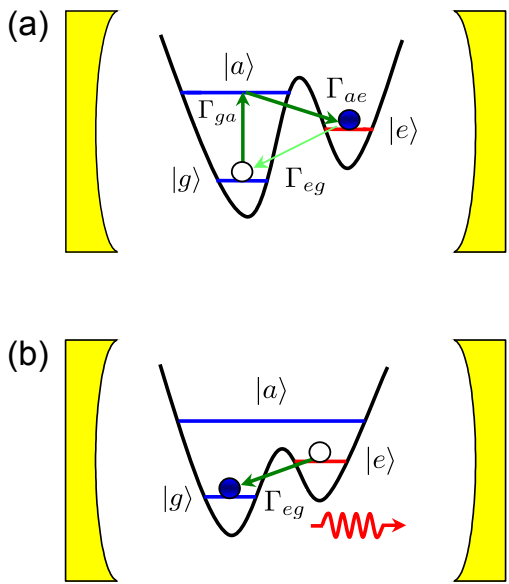} \caption{{\bf Lasing.} {\bf (a)}~State population
inversion (for lasing) between states $|e\rangle$ and $|g\rangle$ in
a three-junction loop at $f>\frac{1}{2}$, where the artificial atom
is quickly pumped from $|g\rangle$ to $|a\rangle$ by a strong
microwave pulse (for example, by a quick Rabi oscillation) and then
decays from $|a\rangle$ to $|e\rangle$ via photon emission. Here the
dipole transition rate from $|e\rangle$ to $|g\rangle$ is small
because of a higher inter-well barrier between them; the dipole
transition rate from $|a\rangle$ to $|e\rangle$ is larger owing to a
lower barrier; and the rate is even larger for the dipole transition
$|g\rangle\rightarrow|a\rangle$ because no potential barrier is
involved in the single well.
{\bf (b)}~Photon emission of the artificial atom to build up a
lasing field, where the inter-well barrier is lowered so as to have
a strong transition rate between $|e\rangle$ and $|g\rangle$ and to
tune the transition $|e\rangle\rightarrow|g\rangle$ into resonance
with the cavity.} \label{fig4}
\end{figure}

With suitable junction parameters, the flux qubit circuit can have
the following dipole transition rates~\cite{maser} when
$f\neq\frac{1}{2}$: $\Gamma_{ga}>\Gamma_{ae}\gg\Gamma_{eg}$, where
the rate $\Gamma_{ij}$ is proportional to $|t_{ij}|^2$, with
$t_{ij}$ being the dipole transition matrix element between states
$|i\rangle$ and $|j\rangle$. Because the transition
$|a\rangle\rightarrow|e\rangle$ can be dominant over
$|e\rangle\rightarrow|g\rangle$, an SPI between states $|e\rangle$
and $|g\rangle$ [Fig.~\ref{fig4}(a)] can be quickly achieved by
pumping the artificial atom (via the transition
$|g\rangle\rightarrow|a\rangle$) using a strong microwave field.
Here the artificial atom can be placed in, e.g., a coplanar
waveguide resonator~\cite{resonator} by sharing a segment of the
circuit loop with the transmission line. In the process of SPI, the
weak-coupling regime is assured by having the flux qubit circuit
off-resonant with respect to the cavity.

While the SPI is being established, the biasing flux can be adjusted
to give a value of $f$ near $\frac{1}{2}$, where the rate
$\Gamma_{eg}$ for the transition $|e\rangle\rightarrow|g\rangle$
becomes large and the cavity is resonant with this transition.
However, the adiabatic condition is not easy to satisfy near this
anticrossing point ($f=\frac{1}{2}$), where the Landau-Zener
transition is strong, so the biasing flux cannot be changed very
fast while approching this point. Fortunately, away from this
anticrossing point, the Landau-Zener transition is weak, so it is
easy to satisfy the adiabatic condition~\cite{maser} and the flux
can be changed very quickly. To take advantage of this property, the
small Josephson junction in the flux-driven loop can be replaced by
a tunable superconducting quantum interference device (SQUID). In
this d.c. SQUID, the magnetic field applied to the loop causes the
critical current to oscillate with period $2\Phi_0$. With the SPI
established at a biasing flux away from $f=\frac{1}{2}$, one can
quickly change the flux in the SQUID loop to lower the inter-well
barrier, so as to both increase the transition rate $\Gamma_{eg}$
and tune this transition $|e\rangle\rightarrow|g\rangle$ into
resonance with the cavity mode [Fig.~\ref{fig4}(b)]. This can yield
a strong coupling between the circuit and the cavity.


To build up a lasing field, in addition to the above two conditions
for quickly establishing SPI and then achieving a strong
circuit-cavity coupling, the cavity used should have a high quality
factor $Q$ ($Q$-factor), i.e., a small leak or decay rate. This can
be implemented using a coplanar waveguide
resonator~\cite{resonator}.

Indeed, lasing was experimentally observed using a Cooper-pair box
in an on-chip cavity~\cite{lasing}. In contrast to the
proposal~\cite{maser} using a flux-driven loop, this
experiment~\cite{lasing} employs a different three-level system:~the
two lowest superconducting states $|g\rangle$ and $|e\rangle$ and a
quasi-particle state $|a\rangle$. The gate voltage is tuned to
$V_g>e/C_g$ (above the degeneracy point), so the state $|2\rangle$
with an extra Cooper pair in the box becomes the ground state
$|g\rangle$ of the artificial atom, and the state $|0\rangle$ with
zero extra Cooper pair in the box is the excited state $|e\rangle$.
Also, the box is connected to a lead via a tunnel barrier. When
driving the box with a voltage across the tunnel barrier, an SPI
between $|e\rangle\equiv|0\rangle$ and $|g\rangle\equiv|2\rangle$ is
achieved, following quasi-particle tunneling
processes~\cite{Armour}.
In \cite{lasing}, lasing was achieved continuously, with emitted
light escaped from one end of the cavity.

\vspace{.4cm}\noindent

{\bf Cooling.}
%
%
%
There are different methods for cooling atoms, including Dopper
cooling, Sisyphus cooling, side-band cooling, subrecoil cooling, and
evaporative cooling. Some of these techniques can be adapted to cool
a solid-state artificial atom. For instance, the Sisypus cooling
technique was used to cool a flux qubit (i.e., a flux-driven
three-junction or four-junction loop)~\cite{Grajcar}.

\begin{figure}
\includegraphics[width=3.4in,bbllx=114,bblly=198,bburx=508,bbury=465]
{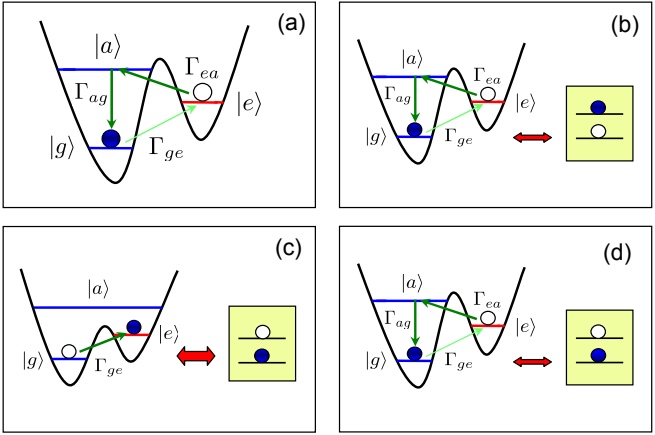} \caption{{\bf Cooling a three-level artificial atom and a
nearby two-level system.} {\bf (a)}~Cooling the three-junction loop
to its ground state $|g\rangle$. While the artificial atom is
thermally excited to $|e\rangle$, one can drive the atom to
$|a\rangle$ by a microwave field. Because of a large transition rate
for $|a\rangle\rightarrow|g\rangle$, the atom can decay quickly from
the unstable state $|a\rangle$ to the ground state, emitting net
energy, extracted from the atom, to the outside environment. The
levels in the double potential well correspond to the lowest three
energy levels of a three-junction loop biased at $f>\frac{1}{2}$.
{\bf (b)}~While the noise source is thermally excited, the
artificial atom is shifted off-resonance to the noise source by
tuning the externally applied flux and also driven to the cooled
state via the inverse process of the state population inversion in
(a). The box to the right of the energy level diagram represents a
two-level fluctuator acting as a noise source. {\bf (c)}~The
inter-well barrier of the artificial atom is lowered by tuning the
externally applied flux so as to have both a strong transition rate
between $|g\rangle$ and $|e\rangle$ and the transition
$|g\rangle\rightarrow|e\rangle$ in resonance with the two-level
system, so as to extract energy from the two-level system. {\bf
(d)}~Shifting the artificial atom off-resonance from the two-level
system and cooling the atom again, with the net energy extracted
from the two-level system emitted to the outside environment.}
\label{fig5}
\end{figure}

Another important advancement~\cite{Valen} is the cooling of a flux
qubit implemented via the inverse process of SPI.
In \cite{Valen}, the temperature of the superconducting qubit was
lowered by up to two orders of magnitude when its surroundings
reached a temperature as low as tens of mK. This cooling of the flux
qubit is achieved when the biasing flux is shifted away from
$f=\frac{1}{2}$, where the dipole transition rates for the lowest
three levels of the flux qubit satisfy the relations
$\Gamma_{ag}>\Gamma_{ea}\gg\Gamma_{ge}$. By optically pumping the
qubit via the transition $|e\rangle\rightarrow|a\rangle$, the qubit
is excited to the high-energy state $|a\rangle$ and then decays to
the ground state by way of the transition
$|a\rangle\rightarrow|g\rangle$ [see Fig.~\ref{fig5}(a)], with a net
energy, extracted from the qubit, emitted to the outside
environment. This experiment also provides fine analogies between
solid-state artificial atoms and natural atoms, as well as showing
how these analogies can inspire new applications.

Although the superconducting qubit in \cite{Valen} was greatly
cooled ($k_BT\ll E_e-E_g$) in experiments, the noise sources
surrounding the qubit were not. So the qubit will quickly return to
the temperature of its environment.
To overcome this difficulty, the superconducting qubit can be
redesigned to increase its controllability by replacing the small
Josephson junction in the flux-driven loop with a tunable
SQUID~\cite{You-PRL}. The cooling process can now be described as
follows [see Figs.~\ref{fig5}(b)-\ref{fig5}(d)]: First, as in the
experiment~\cite{Valen}, the qubit is initially cooled, following
the inverse process of the SPI. Then, the tunable qubit is switched
on for a period of time in order to resonantly interact with the
noise source (e.g., local two-level fluctuators) surrounding the
qubit. This process extracts energy from the noise source to heat
the qubit. Repeating these two processes~\cite{You-PRL}, both the
qubit and its neighboring noise source can be simultaneously cooled.
This will significantly enhance the quantum coherence of the qubit,
because the cooled qubit is then thermally activated only very
slowly to the first excited state.

Recent technical advances allow the fabrication of a nanomechanical
resonator with both a high $Q$-factor and a sufficiently high
frequency, close to the typical frequencies of superconducting
circuits~\cite{Huang-Nature}. This has stimulated researchers to
propose different ways to use superconducting circuits to achieve
the ground-state cooling of coupled nanomechanical
resonators~\cite{Martin,Zhang05,Marquardt,Hauss,You-PRL,Grajcar-PRB}.
Moreover, the experimental cooling of such a resonator, by coupling
it to a superconductor single-electron transistor~\cite{Naik} or to
a microwave-frequency superconducting resonator~\cite{Schwab-10},
has also been reported. When a nanomechanical resonator is cooled to
the ground state~\cite{GS-Cooling}, it provides a good platform for
exploring various quantum phenomena and for observing the
quantum-to-classical transition in such a macroscopic object. This
will give rise to the new subject of quantum acoustics.

\vspace{.4cm}\noindent

\begin{figure}
\includegraphics[width=3.4in,bbllx=68,bblly=418,bburx=555,bbury=752]
{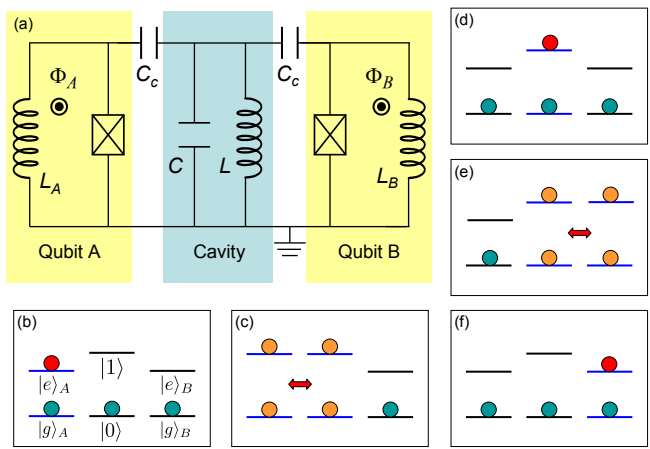} \caption{{\bf Transferring quantum information between
two stationary qubits via a cavity}. {\bf (a)}~Schematic diagram of
two flux-driven phase qubits capacitively coupled by an on-chip
cavity (an $LC$ resonator). {\bf (b)}~Qubit A is prepared in a
superposition state $\alpha|g\rangle_A+\beta|e\rangle_A$, while both
qubit B and the resonator are prepared in their ground states. In
this step, both qubits A and B are off-resonance with the cavity.
{\bf (c)}~Qubit A is shifted into resonance with the resonator, for
a time interval $t_1=\pi/2g_A$, with $\hbar g_A$ being the
interaction energy between qubit A and the resonator. This step maps
the state of qubit A to the superposition state
$\alpha|0\rangle+\beta|1\rangle$ of the resonator, where $|0\rangle$
and $|1\rangle$ are two Fock states of the resonator with zero and
one photon, respectively. {\bf (d)}~Shift qubit A off-resonance with
the resonator again, and store the quantum information in the
resonator for a time duration $t_2$. {\bf (e)}~Shift qubit B in
resonance with the resonator for a time interval $t_3=\pi/2g_B$,
where $\hbar g_B$ is the interaction energy between qubit B and the
resonator. This step maps the state of the resonator to the
superposition state $\alpha|g\rangle_B+\beta|e\rangle_B$ of qubit B.
{\bf (f)}~Shift qubit B off-resonance with the resonator again, and
store the quantum information in qubit B. Note that a high-fidelity
state transfer between qubits A and B can be implemented if both the
relaxation and decoherence of the state are negligibly small during
the above processes.} \label{fig6}
\end{figure}

{\bf Photon generation.} Superconducting qubits have the advantage
of manipulating quantum states in a controllable manner. If these
stationary qubits are spatially separated, one can use single
photons generated in an extended cavity as a quantum bus, similar to
a flying qubit, to implement quantum communication among them [see
Fig.~\ref{fig6}(a)]. Technologically, this requires the generation
of single photons by manipulating a superconducting qubit and the
transfer of information between the superconducting qubits and the
photons. Using an on-chip cavity,
it becomes feasible to achieve this quantum communication process on
a chip.

Recent experiments show that a single-photon source can be achieved
using a superconducting qubit coupled to an on-chip
cavity~\cite{transfer,single,Fock,multi}. When the qubit is prepared
in the excited state $|e\rangle$ by a control pulse, it can decay to
the ground state $|g\rangle$ by emitting one (and only one) photon
in the on-chip cavity; this decay is possible because of the
interaction between the qubit and the cavity. If the cavity was
originally in the vacuum state $|0\rangle$, it now changes to the
single-photon state $|1\rangle$. When the qubit is prepared in an
arbitrary superposition state $\alpha|g\rangle+\beta|e\rangle$, in
an ideal manner, the coupling between the qubit and the cavity can
map the qubit state into a superposition state of zero and one
photon in the cavity: $\alpha|0\rangle+\beta|1\rangle$ [see
Figs.~\ref{fig6}(b)-\ref{fig6}(d)]. Furthermore, the experiment in
\cite{transfer} shows how to transfer the information from a cavity
to a stationary qubit [see Figs.~\ref{fig6}(d)-\ref{fig6}(f)]. These
experiments demonstrate that both single-photon sources and quantum
communication between superconducting qubits can be achieved on a
chip. However, because of the relaxation and decoherence in both
qubit and cavity, after each step for transferring the information
between a stationary qubit and a cavity, the amplitudes $\alpha$ and
$\beta$ of the photon (qubit) state can be different from the
amplitudes $\alpha$ and $\beta$  of the previous qubit (photon)
state. Improving relaxation rates and qubit decoherence times would
allow higher-fidelity state transfers between qubits and cavities.

In addition to single-photon generation, one can also generate, as
proposed in \cite{Law,Liu-EPL}, multi-photon Fock states $|n\rangle$
(i.e., the number states of photons) and arbitrary superposition
states $\sum_n c_n|n\rangle$. Indeed, in a recent
experiment~\cite{Huang-Mart}, the controlled generation of pure Fock
states with up to 15 photons was achieved using a superconducting
phase qubit coupled to a microwave on-chip cavity. Moreover, thanks
to the advantages of both on-chip cavity and tunable superconducting
circuits, complex superpositions of states with different number of
photons were also generated in a controlled and deterministic
manner~\cite{multi}, which is a beautiful experimental realization
of the protocol described in
\cite{Law}. 
Recently, the $N$-photon entangled NOON states,
$|N0\rangle+|0N\rangle$, have also been generated in two
superconducting resonator~\cite{Huang-Mart2}. These experiments
further reveal the quantum behavior of the on-chip cavity and
provide a useful on-demand multi-photon source for future
quantum-technology applications.


\vspace{.4cm}\noindent

{\bf Quantum state tomography.} A crucial step in quantum
information processing is the measurement of the output quantum
states. However, a quantum state cannot be ascertained by a single
quantum measurement. This is because quantum states may comprise
many complementary features which cannot be measured simultaneously
and precisely, owing to uncertainty relations. Nevertheless, all
complementary aspects can in principle be observed by a series of
measurements on a large enough number of identically prepared copies
of the quantum system. Then, one could reconstruct an unknown
quantum state from such a complete set of measurements of system
observables. Such a process of reconstructing quantum states is
called quantum state tomography. Using state tomography, the noisy
channel of the quantum system can also be determined. This procedure
of determining the dynamics of an open quantum system is known as
quantum process tomography.

Tomographic measurements on the quantum states of superconducting
charge qubits, either single or multiple qubits, were proposed in
\cite{Liu-Tomo-PRB}. Recently, there were many experiments on the
quantum state tomography of single superconducting phase
qubits~\cite{Tomo-1,Tomo-2} and of two coupled superconducting
phase~\cite{Tomo-3} and charge~\cite{Tomo-4} qubits. Also, quantum
process tomography was experimentally implemented on
single~\cite{Tomo-5} and two~\cite{Tomo-7} phase qubits. Indeed,
quantum state tomography is an essential tool in qubit-state
measurements, and quantum process tomography can be used to probe
the noise properties
and temporal dynamics
of qubit systems.

\vspace{.8cm}\noindent

{\bf\large Future prospects} \vspace{.3cm}\noindent

With technological advances, superconducting circuits can be used to
test quantum mechanics on a macroscopic scale (see Box~3). Also,
they can be used to demonstrate many novel phenomena in quantum
science. A few examples are listed below.

\vspace{.4cm}\noindent

{\bf Dynamical Casimir effect.} When two mirrors are placed in empty
space, their presence affects the vacuum fluctuations of the
electromagnetic field. Because of the different densities of the
vacuum modes inside and outside of the space between the two
mirrors, a net force on the mirrors can be generated. This effect of
quantum electrodynamics is known as the static Casimir effect.

If the mirrors move, there is also a mismatch between vacuum modes
at different times. It has been predicted
that this may result in the creation of real photons out of vacuum
fluctuations. This dynamical Casimir effect also holds for a single
mirror subject to a nonuniform acceleration in empty space.
Although receiving considerable interest since its theoretical
prediction, there is still no experimental verification of the
dynamical Casimir effect. This is mainly due to the fact that the
rate of photon production is non-negligible only when the mirror
velocity approaches the speed of light, making the use of massive
mirrors very challenging.
A coplanar waveguide terminated by a SQUID was
proposed~\cite{Johansson1} for experimentally observing the
dynamical Casimir effect. Changing the magnetic flux threading the
SQUID loop parametrically modulates the boundary condition of the
waveguide and thereby its effective length. Because there is no
massive mirror moving, the velocity of the effective boundary can
approach the speed of light. Photon production from the vacuum can
thus be made experimentally detectable.


\vspace{.4cm}\noindent

{\bf Coherent population transfer.} Elementary logic gates in
quantum computing networks are usually implemented using
precisely-designed resonant pulses. However, the various
fluctuations and operational imperfections that exist in practice
limit these designs. Also, the difficulty of switching interqubit
couplings on and off strongly limits the precise design of the
required pulses for two-qubit gates. To overcome these difficulties,
Ref.~\cite{Wei-08} proposes an approach to coherently transfer the
populations of qubit states by using Stark-chirped rapid adiabatic
passages.
As in the case of geometric phases, these population transfers are
insensitive to the dynamical evolution times of the qubits, as long
as they are adiabatic. The rapid adiabatic passages of populations
could offer an attractive approach to implementing high-fidelity
single- and two-qubit gates for quantum computing.

The key of these rapid adiabatic passages is how to produce
time-dependent detunings by chirping the qubit levels. For most
natural atomic or molecular systems, where each bound state
possesses a definite parity, the required detuning chirps could be
achieved by making use of the Stark effect via, eg., two-photon
excitations of the qubit levels~\cite{Ran}. The breaking of parity
symmetries in the bound states in superconducting circuits such as
current-biased Josephson junctions provides an
advantage~\cite{Wei-08}, because the desirable detuning chirps can
be produced by single-photon pulses.
Recently, rapid adiabatic passage was achieved for the tranfer of a
single photon in a superconducting circuit~\cite{RAP}.

\vspace{.4cm}\noindent

{\bf Tunable mirrors and interferometers.} Superconducting circuits
can be used for Landau-Zener-St{\" u}ckelberg
interferometry~\cite{LZS}, but they can also be used for other types
of interferometry---including Fano and Fabry-Perot
interferometry~\cite{Zhou-PRL,Zhou-PRA}---by coupling
superconducting qubits to a coplanar waveguide. When injected into
the waveguide, the photons interact with the qubits along the way,
controlled by changing the applied electric and/or magnetic fields
on the qubits. These artificial atoms, working as tunable mirrors,
can change the reflection and transmission coefficients of the
photons confined in the waveguide.

\vspace{.01cm}\noindent
\begin{boxedminipage}[t]{3.4in}

{\bf\large Box~3.~Testing quantum mechanics with macroscopic
superconducting circuits}

\vspace{.5cm}\noindent {\bf\it Bell inequality}

The Bell inequality shows that the predictions of quantum mechanics
can contradict those of local hidden variable theories (see, e.g.,
\cite{Geno}) if one looks at correlations between spatially
separated measurements. It can alternatively be stated that no
physical theory of local hidden variables can reproduce all of the
predictions of quantum mechanics. Tests of the Bell inequality have
been proposed, using superconducting circuits such as
charge~\cite{Wei-PRB} and phase qubits~\cite{Kofman-PRA}. Recently,
the violation of the Bell inequality has been experimentally
verified~\cite{Ans} in phase qubits. Because the Bell inequality is
violated by a quantum mechanical prediction, this experiment
provides strong evidence that these macroscopic superconducting
circuits indeed behave quantum mechanically. Recent experimental
results~\cite{GHZ1,GHZ2} on Greenberger-Horne-Zeilinger states do
not require statistical arguments for a violation of the Bell
inequality to be seen.

\vspace{.3cm}\noindent {\bf\it Leggett-Garg inequality}

Leggett and Garg derived an inequality for a single degree of
freedom undergoing coherent oscillations and being measured at
successive times~\cite{LG}. The Leggett-Garg inequality can be
regarded as a temporal version of Bell's inequality, and it should
be violated by a quantum two-level system. Very recently, this has
been verified experimentally~\cite{Pala} using a voltage-driven box
(i.e., the Cooper-pair box) acting as a quantum two-level system,
showing that the time correlations present at the detector output
violate the inequality.

\vspace{.3cm}\noindent {\bf\it Kochen-Specker theorem}

This theorem elucidates the conflict between quantum mechanics and
noncontextual hidden-variable theories~\cite{Geno}. Noncontextuality
means that the measured value of an observable is independent of the
choice of other co-measurable (commuting) observables that are
measured previously or simultaneously. Quantum mechanics is
contextual, because outcomes depend on the context of measurement.
This theorem is an important complement to Bell's theorem; testing
it can disprove noncontextual hidden-variable theories without
referring to locality. To confirm such a counterintuitive phenomenon
on a macroscopic scale, it was proposed~\cite{Wei-preprint} to use
two charge qubits which are controllably coupled by a two-level
data-bus built from a phase qubit. The analysis~\cite{Wei-preprint}
showed that by performing joint nondestructive quantum measurements
of two distinct qubits, the proposed superconducting circuits could
demonstrate quantum contextuality at a macroscopic level.

\end{boxedminipage}
\vspace{.2cm}\noindent

For a system consisting of a superconducting qubit in an array of
coupled cavities, the photon transmission exhibits a more general
line shape~\cite{Zhou-PRL}, beyond the Breit-Wigner and Fano line
shapes, because of the nonlinear photonic dispersion relation. At a
particular matching condition between the photon wavelength and the
lattice constant~\cite{Zhou-PRL}, the photonic dispersion relation
can become linear and the photon transmission has the Breit-Wigner
line shape, just as in an open transmission line~\cite{Fan-PRL}.
Recently, this phenomenon was observed in a superconducting flux
qubit coupled to an open transmission line~\cite{Oleg-10}. When two
superconducting qubits are placed in an array of coupled cavities,
they can be used as tunable mirrors to form a Fabry-Perot
interferometer~\cite{Zhou-PRA}. Such a controllable on-chip
interferometer is expected to have various applications in quantum
optics.

\vspace{.4cm}\noindent

{\bf Quantum nondemolition measurements.} In a quantum measurement,
a signal observable of a quantum system is measured by detecting the
change in an observable of the detector that is coupled to the
quantum system during the process of measurement. Generally, the
process of measurement will disturb the state of the quantum system
owing to the interplay between the system and the detector. A
quantum nondemolition (QND) measurement does not perturb the
subsequent evolution of the quantum system; this can be achieved by
using a particular type of system-detector coupling that preserves
the eigenstates of the signal observable in the quantum system. In
quantum optics, a QND measurement of the photon number can be
implemented using the optical Kerr effect and a dispersive
atom-field coupling (see, e.g., \cite{Scully}).

The first successful QND measurement on a superconducting qubit was
implemented using the dispersive atom-field coupling
technique~\cite{resonator,Wallraff-PRL-05}. Recent
experiments~\cite{Lupascu,Boulant} show that QND measurements can
also be implemented for a single superconducting qubit by using a
nonlinear resonator as the detector. In \cite{Lupascu}, the detector
was composed of a SQUID shunted with a capacitance, while in
\cite{Boulant} the detector was a bifurcation amplifier, which is a
r.f.-driven Josephson junction working near the dynamical
bifurcation point~\cite{Siddiqi}. Very recently, a fast QND
measurement of a flux qubit was implemented in the weakly-projective
regime by employing a hysteretic d.c. SQUID detector~\cite{Picot}.
A quantum device can have multiple qubits, so QND measurements on
quantum states of multiple qubits (e.g., entangled states) should be
an appealing topic for future investigations.


\vspace{.4cm}\noindent

{\bf Generating squeezed states.} Squeezed states have been
extensively studied in quantum optics, and are now being studied in
condensed matter systems. Owing to their tunable nonlinearity and
low losses in the microwave regime, Josephson-junction
superconducting circuits are promising devices for producing
squeezed states. In superconducting circuits, $LC$ oscillators have
been successfully used for quantum control and readout devices in
conjunction with superconducting qubits.
As parametric transducers (essentially, a radio-frequency
auto-oscillator), superconducting resonant tank circuits have been
used to measure the quantum state of flux qubits~\cite{tank}. When
squeezed states are generated in these resonant tank circuits acting
as quantum-state detectors,
the noise of the detectors can be decreased below the standard
quantum limit. A recent theoretical study~\cite{Zagoskin} shows that
a superconducting parametric transducer can naturally implement this
approach, as it can be used both to produce squeezed states and to
use them in order to minimize quantum fluctuations. An immediate
application of this method would be to suppress the effective noise
temperature of the amplifier connected to the parametric transducer,
at least to the nominal temperature of the cooling chamber.

\vspace{.4cm}\noindent

{\bf Topological phases.} A topologically-protected quantum state
degeneracy cannot be lifted by any local
perturbations~\cite{Nayak-RMP}. It is therefore natural to consider
using topological phases for applications requiring a high degree of
quantum coherence. With superconducting circuits as building blocks,
various artificial lattices can be constructed that possess
interesting topological phases. For instance, it has been proposed
that a triangular Josephson junction array may have a two-fold
degenerate ground state, which could be used for constructing
topologically-protected qubits~\cite{Ioffe}. Recently, an
experiment~\cite{Gladchenko} was implemented for a prototype device
that consisted of twelve physical qubits made of nanoscale Josephson
junctions. Owing to properly tuned quantum fluctuations, this system
was protected against magnetic flux variations well beyond linear
order. This suggests that topologically protected superconducting
qubits are feasible. Also, superconducting circuits were
proposed~\cite{You-PRB-10} as a way to construct the Kitaev
honeycomb model, which requires that the spin (natural or
artificial) at each node of a honeycomb lattice interacts with its
three nearest neighbors through three different types of
interactions~\cite{Kitaev}. Depending on the bond parameters, this
anisotropic spin model supports both Abelian and non-Abelian anyons,
which are particles obeying unusual statistics (they are neither
bosons nor fermions). Its realization would provide exciting
opportunities for experimentally demonstrating anyons.

\vspace{.4cm}\noindent

{\bf Final remarks.} The superconducting circuits that we have
described above contain Josephson junctions that can act as
nonlinear inductors. Using suitably designed superconducting
circuits, it is therefore possible to fabricate field-controlled
nonlinear resonators, which can be used to demonstrate the Kerr
effect (either quadratic electro-optic or quadratic magneto-optic).
If such circuits were used as a Kerr medium, one could carry out a
variety of nonlinear optics experiments, e.g., coupling microwave
photons, implementing quantum gates for photon qubits, and
performing QND measurement. Superconducting circuits could have many
other applications. For instance,
it has been suggested that a coplanar waveguide with the center
conductor replaced by an array of SQUIDs could be used to simulate
Hawking radiation~\cite{Nation}. Indeed, superconducting circuits
have the advantage of enabling the study of complex controllable
quantum dynamics.
This could lead to quantum simulations and on-chip studies of
many-body physics. Numerous new phenomena and applications will
continue to be discovered using superconducting circuits, and these
will play an important part in future quantum technologies.



\vspace{.3cm}\noindent
{\small {\bf Acknowledgments} We thank S. Ashhab for comments on the
manuscript. J.Q.Y. acknowledges partial support from the National
Basic Research Program of China grant No. 2009CB929300, the National
Natural Science Foundation of China grant No. 10625416, the ISTCP
grant No. 2008DFA01930, and the MOE grant No. B06011. F.N.
acknowledges partial support from the Laboratory of Physical
Sciences, National Security Agency, Army Research Office, DARPA,
AFOSR, National Science Foundation grant No. 0726909, JSPS-RFBR
contract No. 09-02-92114, Grant-in-Aid for Scientific Research (S),
MEXT Kakenhi on Quantum Cybernetics, and the JSPS through its FIRST
Program.


\end{document}